# Landau's Nobel Prize in Physics

Mats Larsson (a) , A.V. Balatsky (b,c)

(a) Department of Physics, AlbaNova University Center, Stockholm University, SE-10691 Stockholm, Sweden
(b) Institute for Materials Science, Los Alamos National Laboratory, Los Alamos, NM 87545, USA
(c) Nordita, Center for Quantum Materials, Roslagtullsbacken 23, SE-10691 Stockholm, Sweden

Version May 9, 2016

*Abstract*: Work of Lev Landau had a profound impact on the physics in 20$^{th}$ century. Landau had created the paradigms that had framed the conversations on the outstanding problems in physics for decades. He has laid the foundations for our understanding of quantum matter such as superfluidity, superconductivity and the theory of Fermi Liquid. Here we present sampled Nobel Archive data on the winning nomination that led to the Nobel Prize in Physics in 1962.

Lev Landau [1] was awarded the Nobel Prize in Physics in 1962 *"for his pioneering theories for condensed matter, especially liquid helium"*[2]. On January 7 1962, during Russian Christmas season, Lev Landau had been injured in a car accident outside Moscow. Landau lived another six years; however, he could not again work at same level or activity as a physicist. The car accident was not mentioned in the nominations nor in Nobel Committee deliberations and it is not clear how it affected the nominations and the Nobel Committee considerations. Although it is known that he could not get to Stockholm in December 1962 as a result of illness. As a result, the prize ceremony was held in the Swedish Embassy in Moscow without a Nobel Lecture.

One can speculate that perhaps this lecture would have laid out the vision and perspective of a laureate and would be an important part of the scientific legacy. The undelivered lecture is one of many twists and turns in the life of Lev Landau

With the passing of the fiftieth Noble Prize we have had access to the Nobel Archive and to the nomination records for Landau's Nobel Prize. Here we present a summary of these records. The nomination letters are solicited from the community and used as a means to access the strength of various nominations. These letters have laid the basis for the winning nominations for 1962 award and were written by W. Heisenberg (Max-Planck-institute for Physics and Astrophysics) and by J.Pellam (Caltech). In addition, a letter was sent by N. Bohr, cosigned by A. Bohr, B. Mottelson, C. Møller and L. Rosenfeld. This particular letter did not reach the Nobel Committee in time for the dead line of January 31, 1962, and did not count as a nomination for the 1962 Nobel Prize. However, it is clear from the record of the deliberations of the Nobel Committee that the nomination by N. Bohr was very influential. Heisenberg nominated Landau for the prize earlier, in 1959 and in 1960 as well. Overall Landau was nominated for the Nobel prize in physics from 1954 to 1963 21 times, of which 9 times together with Pyotr Kapitsa (Nobel Prize in Physics 1978) [3]. Since Kapitsa was not nominated in 1962, he could not be considered for a shared prize with Landau in 1962. We present here copies of the 1962 nomination letters for the record.

# MAX-PLANCK-INSTITUT FÜR PHYSIK UND ASTROPHYSIK

**INSTITUT FÜR PHYSIK**

Prof. W. Heisenberg

MÜNCHEN 23, den 10. Jan. 1962
AUMEISTERSTRASSE 6
TELEFON 363201

KVA:s Nobelkommittéer
Inkom den 15. 1 1962

An den
Herrn Präsidenten des
Nobelkomitees für Physik
der Königl. Akademie der Wissenschaften

S t o c k h o l m  50 (Schweden)

Sehr verehrter Herr Präsident!

Für den Nobelpreis für Physik des Jahres 1962 möchte ich noch einmal Herrn L. L a n d a u in Moskau vorschlagen. Wie Sie wissen, hatte ich Herrn Landau schon früher vorgeschlagen und als Begründung seine Untersuchungen über die Quantentheorie des Diamagnetismus, des superfluiden Heliums und insbesondere seine Arbeiten zur Quantenfeldtheorie genannt. Ich habe auch erwähnt, daß man bei Landau, der ja ein sehr vielseitiger theoretischer Physiker ist, vielleicht nicht eine einzige besonders glanzvolle Entdeckung hervorheben kann, daß aber sein Gesamtwerk so bedeutend ist, daß man die Verleihung des Nobelpreises an ihn in jeder Weise rechtfertigen kann. Auch früher sind die Statuten der Stiftung ja schon in dieser Weise ausgelegt worden. Ich denke etwa an die Verleihung des Preises an Debye, Bothe und Born. Nachdem in den vergangenen Jahren verschiedene russische Kollegen den Nobelpreis in Empfang genommen haben, kann man ja wohl auch annehmen, daß aus einer solchen Verleihung für Landau sich keine Schwierigkeiten ergeben würden.

Mit den besten Empfehlungen
Ihr sehr ergebener

W. Heisenberg

Fig. 1 . Nomination letter by W. Heisenberg. Reproduced with permission from the Nobel Archive of the Royal Swedish Academy of Sciences.

January 26, 1962

The Nobel Committee for Physics
Stockholm 50
Sweden

Dear Sirs:

Thank you for the invitation to nominate a candidate for the Nobel Prize for Physics for 1962.

My nomination is for Lev Davidovich Landau of the Institute for Physical Problems, Moscow.

The justification is based on a series of momentous contributions to the theory of liquid helium over a period of some twenty years. Specifically, the first of this series:

> The Theory of Superfluidity of Helium II, by L. Landau, Journal of Physics USSR, V (1), 71 (1941)

set up the quantum mechanics for liquids or "quantum hydrodynamics". Specifically the idea of the energy gap $\Delta$ was introduced for "roton" excitations and the commutation rules for circulation operators introduced. The special rotation conditions requiring discretely (as opposed to continuously) varying circulation properties were expounded and constituted the foundation for "quantized rotations".

This early treatment of the liquid helium II problem was followed by the Letter in 1949:

> On the Theory of Superfluidity, L. Landau, Physical Review (Letter) 75, 884 (1949).

in which Landau presented the form of the energy spectrum ($\varepsilon$ vs p) for the elementary excitations. This spectrum, which both encompasses and connects the phonon region and the roton region, has since been amply verified experimentally in various ways.

About the same time Landau, in collaboration with I.M. Khalatnikov, published two major papers:

> The Theory of the Viscosity of Helium II.
> I. Collisions of Elementary Excitations in Helium II
> II. Calculation of the Viscosity Coefficient
> by L.D. Landau and I.M. Khalatnikov, J. Exp. Theor. Physics, USSR 19, 637 and 709, respectively (1949).

In these the energy spectrum for excitations was applied to second order processes to provide scattering cross-section calculations for phonon-phonon, phonon-roton, and roton-roton collisions. Extension of these results to second sound attenuation by Khalatnikov and subsequent experimental verification has confirmed Landau's theory also in its second order application.

In recent years Landau has turned his attention to the properties of the light isotope, helium 3. Specifically in connection with liquid helium 3, the following publications:

>The Theory of a Fermi Liquid, L.D. Landau, JETP USSR 30, 1058 (1956); JETP (US) 3, 920 (1956-7)
>Oscillations in a Fermi Liquid, L.D. Landau, JETP USSR 32, 59 (1959); JETP (US) 5, 101 (1957)
>Contributions to the Theory of the Fermi Liquid, L.D. Landau, JETP USSR 35, 97 (1958); JETP (US) 8, 70 (1959)

treat the enticing helium 3 problem. Landau has predicted a new form of wave propagation in liquid helium 3 which he has named "zero sound". Experimental state of the art has not provided opportunity for testing this latest prediction.

Thus Landau's contributions to the theory of liquid helium have extended over a period of nearly two decades and are still pointing the way to experimentalists in this exerting field.

Unfortunately I have no reprints of any of the above publications. In case the Committee does not already possess copies of any (or all) of these items please notify me and I will have photo-copies (in English) prepared and forwarded. In connection with Landau's over-all impact on the field of the field of low temperature physics I enclose reprints of a talk presented on the occasion of his winning the Second Fritz London Award.

                                        Sincerely yours,

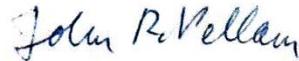

                                        John R. Pellam
                                        Professor of Physics

Enclosures: 3

**Fig.2.** Nomination letter by J. Pellam. These letters laid the basis for the winning nomination for Landau Nobel Prize awarded in 1962. Reproduced with permission from the Nobel Archive of the Royal Swedish Academy of Sciences.

Heisenberg did mention contributions of Landau to development of quantum theory, theory of diamagnetism and superfluidity of He II. In his nomination he points to the significant contributions of Landau, but also describes the possible difficulty in singling out one discovery of Landau that could motivate a Nobel Prize in Physics (according to the will of Alfred Nobel, the physics prize should be awarded for a "discovery" or "invention"). Heisenberg notes that also at earlier occasions the rules have allowed Nobel Prize awards for overall impact and achievements and mentions Debye (Nobel Prize in Chemistry in 1936, Bothe and Born (divided Nobel Prize in Physics in 1954). Heisenberg did not see any issues in justifying a Nobel Prize in Physics to Landau considering his many important contributions to physics.

J. Pellam's nomination was focused on the work of Landau in theory of quantum fluids. The two major aspects of Landau work were mentioned: theory of quantum mechanics of "quantum hydrodynamics" of the Bose liquid and theory of fermi liquid. The letter points the pioneering work of Landau on the superfluidity of HeII starting with the first paper by Landau "The superfluidity of He II", Journal of Physics USSR, v 1, p 71 (1941) [4] that was followed by the subsequent papers (single authored and with I. Khalatnikov) in 1949. The subsequent discovery of roton spectrum and role of roton and phonon scattering in the second sound attenuation have provided an explicit experimental validation of the superfluid theory. In the second part of the letter J. Pellam highlights the foundational work on Fermi liquid theory. Starting with the first paper on "The Theory of a Fermi Liquid" L. Landau, JETP USSR, v 30, p 1058 (1956) [5] and later covering the subsequent work, the author emphasizes the predictions of the zero sound as a new mode of fermi liquids that was earlier unobserved. Pellam summarized the sustained impact of Landau's work on the field of quantum liquids over the span of two decades and pointing experimentalists the way in this field.

A letter was also sent by N. Bohr and A. Bohr, B. Mottelson, C. Møller and L. Rosenfeld. It also demonstrates the wide ranging influence of the work of Landau over the years. Bohr and coauthors made an emphasis not on the particular contribution of Landau to physics but on the rare analytical ability to create the framework that allowed one focus on the most important aspects of physics at the time. They compared Landau to Lord Rayleigh and Lorentz. Unfortunately the letter, while sent in January 30 1962 from Copenhagen did not arrive in time to the Nobel Committee in Stockholm to be taken as a valid nomination by the committee for the year 1962. It was, however, clearly influential during the deliberation process by the Committee.

The last sentence is well worth translating into English (translation by ML):

"As is apparent from reasons put forward in support of our proposal, we have not intended to emphasize any particular of Landau's many important contributions to physics, but much more to put weight on the fact that he has in publication after publication enriched the physics of our time in way which is difficult to find a parallel to, and which reminds of how scientists such as Lord Rayleigh and Lorentz with their extraordinary analytical abilities managed to clarify those problems that confronted science during their time."

The Nobel Committee appointed one of its members, Ivar Waller, to prepare a report about Landau. He writes on September 20, 1962 (translated from Swedish by ML):

"Landau's importance has above all been that he with extraordinary intuition opened and paved the way for the continuing development of research in several areas of condensed matter and in this way, to an unusually high degree, inspired and stimulated this research".

- 3 -

sound", som snart efter fandt sin bekræftelse ved Peshkovs eksperimenter.

Som det vil fremgå af den i denne skrivelse anførte begrundelse af vort forslag, har vi ikke ment at burde fremhæve noget enkelt af Landaus mange betydningsfulde bidrag til den fysiske videnskab, men langt snarere at lægge vægt på, at han skridt for skridt har beriget vor tids fysik på en måde, til hvilken et sidestykke vanskeligt kan findes, og som minder om, hvordan forskere som Lord Rayleigh og Lorentz med deres enestående analytiske evner formåede at klarlægge de problemer, som videnskaben i deres tid stilledes overfor.

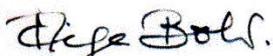
Aage Bohr

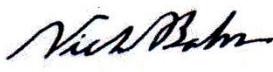
Niels Bohr

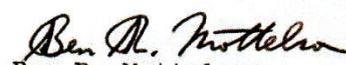
Ben R. Mottelson

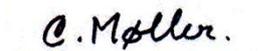
Christian Møller

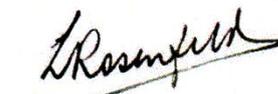
Léon Rosenfeld

**Fig. 3.** The end of the letter of nomination by Niels Bohr (and four younger physicists in Copenhagen). The letter is written in Danish. One can assume that Niels Bohr was more comfortable writing this letter in his native Danish language than in English. He was of course aware that the Swedish Nobel Committee had no problems in understanding Danish. It is not known by the authors why the letter, dated January 30, 1962, did not arrive to the Nobel Committee until February 28 1962. Reproduced with permission from the Nobel Archive of the Royal Swedish Academy of Sciences.

It is interesting to note that the groundbreaking work of Ginzburg and Landau in 1950, which was acknowledged in V.L. Ginzburg Nobel prize in 2003, jointly with A. Abrikosov and A. Legget [6], was not mentioned in the written documentation of the 1962 Nobel Prize in Physics

Impact of Landau's body of work is now fully appreciated [7,8]. Nobel prize was a pinnacle of his career cut short by the car accident. We also mention that in addition to the profound body of work in physics Landau has created a school of physics that has an international recognition.  There are other sides to the Landau impact on physics both in USSR and internationally. Courses on Theoretical Physics and theoretical minimum exams are but one example. We also mention Landau Institute for Theoretical Physics that was founded in 1964. Institute has created the structure that allowed his colleagues, followers and students to continue the style of theoretical physics advocated by Landau.

.

Work was supported by US DOE BES, Aspen Center for Physics, KAW and Swedish Research Council. We are thankful to access to the Nobel Archive of the Royal Swedish Academy of Sciences, and to Karl Grandin, head of the archive, for valuable comments. AVB is grateful to D. Balatsky for comments and critique.